\documentstyle[preprint,eqsecnum,aps,epsfig]{revtex} 

\newcommand{\beq}{\begin{equation}}
\newcommand{\eeq}{\end{equation}}
\newcommand{\nin}{\noindent}

\begin{document}

\title{Quasilinearization Approach to Nonlinear Problems in Physics with
Application to Nonlinear ODEs} 

\author{V.~B.\ Mandelzweig$\,^{1}$\footnote{Electronic mail:
        victor@helium.phys.huji.ac.il} and F.~Tabakin$\,^{2}$
       \footnote{Electronic mail: tabakin@pitt.edu}}

\address{$^1$~Racah Institute of Physics, Hebrew University,
         Jerusalem 91904, Israel\\$^2$~Department of Physics and 
         Astronomy, University of Pittsburgh, Pittsburgh, PA 15260,USA}

\maketitle

\begin{abstract}
\noindent
The general conditions under which the quadratic, uniform and monotonic
 convergence in the quasilinearization method of solving nonlinear ordinary  
 differential equations could be proved are formulated and elaborated. The 
 generalization of the proof to partial differential equations is straight forward. The method, whose  mathematical basis in physics was discussed 
 recently by one of the present  authors (VBM), approximates the solution
 of a nonlinear differential equation by treating the nonlinear terms as a perturbation about the linear ones, and unlike perturbation theories is not based on the existence of some kind of a small parameter.
 
It is shown that the quasilinearization method  gives excellent results when 
applied to different nonlinear ordinary differential equations in physics, 
such as the Blasius, Duffing, Lane-Emden and Thomas-Fermi equations. The
 first few quasilinear iterations already provide extremely accurate and
numerically stable answers.

\noindent

\end{abstract}

\pagebreak

\section{Introduction}

In a series of recent papers,~\cite{VBM,KM00} the possibility of
applying a very powerful approximation technique
called the quasilinearization method (QLM) to physical problems 
has been discussed.
 The QLM  is designed to confront the nonlinear aspects of physical
 processes. The method, whose iterations are constructed to yield 
 rapid convergence and often monotonicity, was originally introduced
 forty years ago by Bellman and Kalaba \cite{K59,BK65} as a 
 generalization of the Newton-Raphson method \cite{CB,RR}  to solve
individual or systems of nonlinear ordinary and partial differential 
equations. Modern developments and
applications of the QLM to different fields are given in a monograph
\cite{LV98}. 

However, the QLM was never systematically studied or extensively
applied in physics, although references to it can be found in well
known monographs \cite{C67,B68} dealing with the variable phase approach to
potential scattering, as well as in a few scattered research papers
\cite{AIC96,J88,RV87,HR83}. The reason for the sparse use of the QLM in Physics
 is that the convergence of the method
has been proven only under rather restrictive conditions \cite{K59,BK65},
which generally are not fulfilled in physical applications. Recently,
though, it was shown \cite{VBM} by one of the present authors (VBM) that a
different proof of the convergence can be provided which we will generalize
 and elaborate here so that the
applicability of the method is extended to incorporate realistic physical
conditions of forces defined on infinite intervals with possible singularities
 at certain points.

In the first paper of the series~\cite{VBM},  the quasilinearization 
approach was applied to the nonlinear Calogero equation in a variable
 phase approach to quantum mechanics and the
results were compared with those of perturbation theory and the exact solutions. 
It was found analytically and by
examples that the $n$-th approximation of  the QLM exactly sums $2^{n}-1$ terms
of perturbation theory. In addition, a similar number of terms is
reproduced approximately. The number of the exactly reproduced perturbation
terms thus doubles with each subsequent QLM approximation, and reaches, for
example, 127 terms in the 6-th QLM approximation, 8191 terms in the 12-th
QLM approximation, and so on.

The computational approach in the work \cite{VBM} was mostly
analytical, and therefore one was able to compute only two to three QLM
iterations, mainly for power potentials. Only in the case of the $1/r^2$
potential, could the calculation of QLM iterations be done analytically for
any $n$. 

The goal of the next work~\cite{KM00}  was, by dropping the restriction of
analytical computation, to calculate higher iterations as well as to extend
the analysis to non-power potentials, in order to  better assess 
the applicability of the method and of its numerical stability and the
convergence pattern of the QLM iterations. It was shown that 
 the first few iterations already provide very accurate and
numerically stable answers for small and intermediate values of the coupling
constant and that the number of iterations necessary to reach a given
 precision only moderately increases for larger values of the coupling.
 The method provided accurate and
stable answers for any coupling strengths, including for super singular
potentials for which each term of the perturbation theory diverges and the
perturbation expansion does not exist even for a very small coupling.

The quasilinearization approach is applicable to a general nonlinear
ordinary or partial $n$~-th order differential equation in $N$-dimensional
space. In this paper, we consider the case of nonlinear
ordinary differential equations in one variable which,  unlike
 the nonlinear Calogero equation\cite{C67}
 considered in references\cite{VBM,KM00}, contain not only quadratic nonlinear
terms but various other forms of nonlinearity and not only a first, 
but also higher derivatives. Namely, we apply it to
 a panopoly of well-known and difficult nonlinear ordinary first, second
 and third order differential
 equations and show that again with just a small number of iterations one can
 obtained fast convergent and uniformly excellent and stable numerical
 results.

The paper is arranged as follows: in the second chapter we present the
main features of the quasilinearization approach, while in the  third
 chapter we consider, as a  warm-up exercise, a simple first-order 
 differential
 equation with a nonlinear $n$-th power term and compare its exact
analytic solution with the perturbation theory and with the QLM iterations
 in order to demonstrate the main features of  the quasilinearization
 approach. In the next four chapters,  we apply our method to four well known 
 nonlinear ordinary second and third order differential equations, namely to
 the Lane-Emden, Thomas-Fermi, Duffing, and Blasius equations,  respectively.
 These equations have been extensively studied in the literature. 
 For a sample of recent papers see refs. \cite{Liao,Tsai,Bin,Wang,Majid,GH00,Pert,BMPS}  and references
therein.

The
results, convergence patterns, numerical stability, advantages of the method
 and its possible future applications
are discussed in the final chapter.

\section{The quasilinearization method (QLM)}

The aim of the quasilinearization method (QLM) of Bellman and 
Kalaba \cite{K59,BK65,LV98} based on 
the Newton-Raphson method \cite{CB,RR} is to solve a nonlinear $n$-th order ordinary
 or partial differential equation in $N$ dimensions as a limit of a
 sequence of linear differential equations. This
goal is easily understandable since there is no useful
technique for obtaining the general solution of a nonlinear
 equation in terms of
a finite set of particular solutions, in contrast to 
a linear equation which can often be solved analytically or
numerically in a convenient fashion using superposition.
 In addition, the QL sequence should be
constructed to assure quadratic convergence and, if
possible, monotonicity.

As we have mentioned in the Introduction, we will follow here the derivation
 outlined in ref. \cite{VBM}, which is not based, unlike the derivations
 in refs.  \cite{K59,BK65}, on a smallness of the interval and on the boundness
of the nonlinear term and its functional derivatives, the conditions which
usually are not fulfilled in physical applications.

For  simplicity,  we limit our discussion to 
 nonlinear ordinary differential equation in one variable on
the interval $[0,b],$ which could be infinite:
\beq
\L^{(n)}u(x)=f(u(x),u^{(1)}(x),....u^{(n-1)}(x), x),\  \label{eq:uxf}
\eeq
\nin with $n$ boundary conditions
\beq
 g_k(u(0),u^{(1)}(0),....,u^{(n-1)}(0))=0 ,\; k=1,...l \label{eq:bounconl}
\eeq
\nin and
\beq 
 g_k(u(b),u^{(1)}(b),....,u^{(n-1)}(b))=0 ,\;  k=l+1,..., n .
\label{eq:bounconr}
\eeq
\nin Here $L^{(n)}$ is linear $n$-th order ordinary differential operator and $f$ and
$g_1,g_2,.....,g_n$ are nonlinear functions of $u(x)$ and its $n-1$ derivatives
 $u^{(s)}(x),s=1,...n-1$. The more general case of partial differential 
equations in  N-dimensional space could be considered in exactly the
 same fashion by changing the definition of $L^{(n)}$ to be a linear
 $n$-th order differential operator in partial derivatives and $x$ to be
 an N-dimensional coordinate array.

The QLM prescription \cite{VBM,K59,BK65} determines the $r+1$-th iterative
approximation $u_{r+1}(x)$ to the solution of Eq.\ (\ref{eq:uxf}) as a
solution of the linear differential equation

\begin{eqnarray}
L^{(n)}u_{r+1}(x)=f(u_r(x),u^{(1)}_r(x),.....,u^{(n-1)}_r(x),x) \nonumber \\ +
\sum_{s=0}^{n-1}  \bigl(u^{(s)}_{r+1}(x)-u^{(s)}_r(x)\bigr)\,
f_{u^{(s)}}(u_r(x),u^{(1)}_r(x),.....,u^{(n-1)}_r(x),x),   \label{eq:difrec}
\end{eqnarray} where $u^{(0)}_r(x)=u_r(x)$, with linearized two-point boundary conditions

\beq
\sum_{s=0}^{n-1}  \bigl(u^{(s)}_{r+1}(0)-u^{(s)}_r(0)\bigr)\,
g_{ku^{(s)}}(u_r(0),u^{(1)}_r(0),.....,u^{(n-1)}_r(0),0)=0,\;\ k=1,...l 
\label{eq:bounconleft}
\eeq

\nin
and

\beq
\sum_{s=0}^{n-1}  \bigl(u^{(s)}_{r+1}(b)-u^{(s)}_r(b)\bigr)\,
g_{ku^{(s)}}(u_r(b),u^{(1)}_r(b),.....,u^{(n-1)}_r(b),b)=0,\;\ k=l+1,..., n .
\label{eq:bounconright}
\eeq

\nin
Here the functions $f_{u^{(s)}}={\partial{f}}/{\partial{u^{(s)}}}$ and  
 $g_{ku^{(s)}}={\partial{g_k}}/{\partial{u^{(s)}}},\  s=0,1,...,n-1$ are
functional derivatives of the functionals
 $f(u(x),u^{(1)}(x),....u^{(n-1)}(x), x)$
 and $g_k(u(x),u^{(1)}(x),....u^{(n-1)}(x), x)$, respectively.~\footnote{For
example, in case of a simple nonlinear boundary condition
 $u'(b)u(b)=c$  where c
 is a constant, one has $g(r) \equiv g(u(r),u'(r),r)=
u'(r)u(r)$ so that $g_u=u'(r)$ and $g_{u'}=u(r)$.
 The linearized boundary condition 
\ref{eq:bounconright} has a form $(u_{r+1}(b)-u_r(b))u_r'(b)+
(u_{r+1}'(b)-u_r'(b))u(b)=0$ or $(u_{r+1}(b)u_r(b))'=(u_{r}(b)u_r(b))'$ so
 the nonlinear boundary condition for the initial guess  $u_0(b)u_0'(b)=c$ will be
 propagated to the linear boundary condition for the next iterations.}.

The zeroth approximation $u_{0}(x)$ is chosen from mathematical or physical
considerations. 

To prove that the above procedure yields a quadratic  and often monotonic 
convergence to the solution of
 Eq.~\ref{eq:uxf} with the boundary conditions \ref{eq:bounconl}
 and \ref{eq:bounconr}, we follow reference~\cite{VBM} and consider 
 a differential equation for the difference $\delta u_{r+1}(x)\equiv u_{r+1}(x)-u_r(x)$
 between two subsequent iterations:

\begin{eqnarray}
L^{(n)}\delta u_{r+1}(x)&=& [f(u_r(x),u^{(1)}_r(x),.....,u^{(n-1)}_r(x),x)- 
f(u_{r-1}(x),u^{(1)}_{r-1}(x),.....,u^{(n-1)}_{r-1}(x),x)] \nonumber \\
&+& \sum_{s=0}^{n-1} [\delta u^{(s)}_{r+1}(x)\, 
f_{u^{(s)}}(u_r(x),u^{(1)}_r(x),.....,u^{(n-1)}_r(x),x) \nonumber \\
&-&\delta u^{(s)}_r(x)\,
f_{u^{(s)}}{u_(r-1}(x),u^{(1)}_{r-1}(x),.....,u^{(n-1)}_{r-1}(x),x)].  
\label{eq:difdif}
\end{eqnarray}

\nin
The boundary conditions are similarly given by the
 difference of Eqs. \ref{eq:bounconleft}
 and \ref{eq:bounconright} for two subsequent iterations:
\begin{eqnarray}
\sum_{s=0}^{n-1} [\delta u^{(s)}_{r+1}(0)\ g_{ku^{(s)}}(u_r(0),u^{(1)}_r(0),...
..,u^{(n-1)}_r(0),0) \nonumber \\
-\delta u^{(s)}_r(0)\ g_{ku^{(s)}}(u_{r-1}(0),u^{(1)}_{r-1}(0),.....,u^{(n-1)}_{r-1}
(0),0)]&=&0,
\nonumber \\ k&=&1,...l 
\label{eq:boundifleft}
\end{eqnarray}

\nin

and

\begin{eqnarray}
\sum_{s=0}^{n-1} [\delta u^{(s)}_{r+1}(b)\ g_{ku^{(s)}}(u_r(b),u^{(1)}_r(b),....
.,u^{(n-1)}_r(b),b) 
\nonumber \\
-\delta u^{(s)}_r(b)\ g_{ku^{(s)}}(u_{r-1}(b),u^{(1)}_{r-1}(b),.....,u^{(n-1)}_
{r-1}(b),b)]&=&0,
\nonumber \\ k&=&l+1,...n .
\label{eq:boundifright}
\end{eqnarray}

In view of the mean value theorem\cite{VOL}

\begin{eqnarray}
&&f(u_r(x),u^{(1)}_r(x),.....,u^{(n-1)}_r(x),x)- 
f(u_{r-1}(x),u^{(1)}_{r-1}(x),.....,u^{(n-1)}_{r-1}(x),x)= \nonumber \\
&&\sum_{s=0}^{n-1} \delta u^{(s)}_r(x)\  
f_{u^{(s)}}(u_{r-1}(x),u^{(1)}_{r-1}(x),.....,u^{(n-1)}_{r-1}(x),x)+ \nonumber \\
&&\frac{1}{2}\sum_{s,t=0}^{n-1} \delta u^{(s)}_r(x)\,\delta u^{(t)}_r(x)\  
f_{u^{(s)}u^{(t)}}(\bar{u}_{r-1}(x),\bar{u}^{(1)}_{r-1}(x),.....,\bar{u}^{(n-1)}
_{r-1}(x),x) ,
\label{eq:meanval}
\end{eqnarray}

\nin
where $\bar{u}^{(s)}_{r-1}(x)$ lies between $u^{(s)}_{r}(x)$ and  $u^{(s)}_{r-1}(x)$. Now 
 Eq. \ref{eq:difdif} can be written as 

\begin{eqnarray}
L^{(n)}\delta u_{r+1}(x)-\sum_{s=0}^{n-1} \delta u^{(s)}_{r+1}(x)\, f_{u^{(s)}}(u_r(x),
u^{(1)}_
r(x),.....,u^{(n-1)}_r(x),x)= \nonumber \\
\frac{1}{2}\sum_{s,t=0}^{n-1} \delta u^{(s)}_r(x)\delta u^{(t)}_r(x)\,  
f_{u^{(s)}u^{(t)}}(\bar{u}_{r-1}(x),\bar{u}^{(1)}_{r-1}(x),.....,\bar{u}^{(n-1)}
_{r-1}(x),x) . 
\label{eq:difmean}
\end{eqnarray}
\nin
Denoting $G_r^{(n)}(x,y)$ as the Greens function,  which is the inverse of the 
following differential operator
and incorporates linearized boundary
 conditions \ref{eq:bounconleft} and \ref{eq:bounconright},

\begin{eqnarray}
\tilde{L}^{(n)}=L^{(n)}-\sum_{s=0}^{n-1}f_{u^{(s)}}(u_r(x),u^{(1)}_
r(x),.....,u^{(n-1)}_r(x),x)\ \frac{d^s}{{dx}^s}\; , 
\end{eqnarray}

\nin
one can express the solution for the difference function $\delta u_{r+1}$
 as
\begin{eqnarray}
\delta u_{r+1}(x)=&& \nonumber \\
\frac{1}{2}\int_{0}^{b} G_r^{(n)}(x,y)&&\sum_{s,t=0}^{n-1} \delta u^{(s)}_r(y)
\delta u^{(t)}_r(y)\,  
f_{u^{(s)}u^{(t)}}(\bar{u}_{r-1}(y),\bar{u}^{(1)}_{r-1}(y),.....,\bar{u}^{(n-1)}_{r-1}(y),y)\,  dy .
\label{eq:quadr}
\end{eqnarray} 
\nin
The functions $\delta u^{(s)}_r(y)\delta u^{(t)}_r(y)$ could be taken outside
 of the sign of the integral at some point $y=\bar{x}$ belonging to the interval,
 so one obtains

\begin{eqnarray}
\delta u_{r+1}(x)=
\frac{1}{2}\sum_{s,t=0}^{n-1} \delta u^{(s)}_r(\bar{x})
\delta u^{(t)}_r(\bar{x}) M_{st}(x) .
\label{eq:quadrill}
\end{eqnarray} 
\nin
where $M_{sr}(x)$  equals

\begin{eqnarray}  
M_{st}(x)=\int_{0}^{b} G_r^{(n)}(x,y)
f_{u^{(s)}u^{(t)}}(\bar{u}_{r-1}(y),\bar{u}^{(1)}_{r-1}(y),.....,
\bar{u}^{(n-1)}_{r-1}(y),y) dy
\label{eq:quadrilli}
\end{eqnarray}
\nin

If $M_{st}(x)$ is a strictly positive (negative) matrix for all $x$
 in the interval, then $\delta u_{r+1}(x)$
will be positive (negative), and the monotonic convergence from below (above)
 results.

Obviously, from Eq. \ref{eq:quadr} follows 
\begin{eqnarray}
|\delta u_{r+1}(x)|\leq \it{k_r}(x)||\delta u_r||^2  \label{eq:quadrat}
\end{eqnarray} 

\nin
where $\it{k}_r$ is given by

\begin{eqnarray} 
\it{k_r}(x)=\frac{1}{2}&&\int_{0}^{b} |G_r^{(n)}(x,y)|\sum_{s,t=0}^{n-1}
|f_{u^{(s)}u^{(t)}}(\bar{u}_{r-1}(y),\bar{u}^{(1)}_{r-1}(y),.....,\bar{u}
^{(n-1)}_{r-1}(y),y)| dy 
\label{eq:quadralast}
\end{eqnarray}
\nin
and $||\delta u_r||$ is a maximal value of any of $|\delta \bar{u}^{(s)}_r|$ on
 the interval (0,b).

Since Eq. \ref{eq:quadrat} is correct for any $x$ on the interval (0,b), it
 is correct also for some
$x=\bar{x}$ where $|\delta u_{r+1}(x)|$ reaches its maximum value  $||\delta
 u_{r+1}||$.
One therefore has 

\begin{eqnarray}
||\delta u_{r+1}|||\leq \it{k_r}(\bar{x})||\delta u_r||^2 \label{eq:quadro}
\end{eqnarray}

\nin
Assuming the boundness of the integrand in expression
\ref{eq:quadralast}, that is the existence of the bounding function $F(x)$
 such that
integrand at  $x=\bar{x}$ and at any $y$ is less or equal to $F(y)$, one
 finally has

\begin{eqnarray}
||\delta u_{r+1}|||\leq \it{k}||\delta u_r||^2 \;,\label{eq:quadrom}
\end{eqnarray}

\nin
where 
\begin{eqnarray}
\it{k}=\int_{0}^{b}F(x) dx\;. \label{eq:quadrol}
\end{eqnarray}

The linearized boundary conditions \ref{eq:bounconleft} and \ref{eq:bounconright}
 are obtained from exact boundary conditions \ref{eq:bounconl} and \ref{eq:bounconr}
by using the mean value theorem Eq. \ref{eq:meanval} and neglecting the quadratic
 terms, so that
 the error in using linearized boundary conditions vis-a-vis the exact ones is quadratic
 in the
 difference between the exact and linearized solutions. The maximum difference between
  boundary conditions 
\ref{eq:bounconleft} and \ref{eq:bounconright}
corresponding to two subsequent quasilinear iterations is therefore quadratic in
 $||\delta u_r||$. 
 In view of this result and of Eq. \ref{eq:quadrom},  the difference between the
 subsequent iterative solutions
 of Eq.\ref{eq:difrec} with boundary conditions \ref{eq:bounconleft} and
 \ref{eq:bounconright}
 decreases quadratically with each iteration. In a similar way, one can show
 \cite{VBM} 
that the difference $\Delta u_{r+1}(x)=u(x)-u_r(x)$ between the exact solution and the
 $r$-th iteration is decreasing quadratically as well:
\begin{eqnarray}
||\Delta u_{r+1}|||\leq \it{k}||\Delta u_r||^2 \;.\label{eq:quadroro}
\end{eqnarray}

 A simple induction of Eq. \ref{eq:quadrom}  shows \cite{BK65} that
${\delta}u_{n+1}(x)$ for an arbitrary $l < r$ satisfies the inequality

\beq
{\parallel}\delta u_{r+1}{\parallel}\leq 
  (\it{k}{\parallel}\delta u_{l+1}{\parallel})^{2^{r-l}}/\it{k}, \label{eq:unum}
\eeq

\nin
or, for $l=0$, we can relate the $n+1$~th order result to the 1st iterate by

\beq
{\parallel}\delta u_{n+1}{\parallel}\leq 
((\it{k}{\parallel}\delta u_{1}{\parallel})^{2^n}/k. \label{eq:unun1}
\eeq

\nin
The convergence depends therefore on the quantity
$q_1=\it{k}{\parallel}{u_1-u_0}{\parallel}$, where, as we have mentioned
 earlier, the zeroth iteration
$u_{0}(x)$ is chosen from physical
and mathematical considerations. Usually it is advantageous 
(see discussion below) that $u_{0}(x)$ would satisfy at least one of the boundary conditions.
 From Eq.\ (\ref{eq:unum}) it
follows, however,  that for convergence it is sufficient that just one of the
quantities $q_m=\it{k}{\parallel}\delta u_{m}{\parallel}$ is small enough.
Consequently, one can always hope \cite{BK65} that even if the first
convergent coefficient $q_1$ is large, a well chosen initial approximation
$u_0$ results in the smallness of at least one of the convergence
coefficients $q_m,\ m>1$, which then enables a rapid convergence of the iteration
series for $r > m$.
It is important to stress that in view of the quadratic convergence of the
 QLM method, the
difference $||\Delta u_{r+1}||$ between
 the exact solution and the QLM iteration always converges to zero
 if the difference $\delta u_{r+1}(x)$ between two subsequent QLM iterations
 becomes infinitesimally small.

 Indeed, if $\delta u_r(x)$ is close to zero, it means, since $\delta
 u_{r+1}(x)=\Delta u_r(x)-\Delta u_{r+1}(x)$  that $\Delta u_r(x)=
\Delta u_{r+1}(x)$
 or $Q_r=Q_{r+1}$
 where $Q_r=\it{k} ||\Delta u_r||$. When one assumes the possibility that
  $ Q_r $ and $ Q_{r+1} $  could be  not small, one could conclude
 that the iteration process ``stagnates'', which means convergence to the
 wrong answer or no convergence at all. 

  However, such a
 conclusion is wrong since
 Eq. \ref{eq:quadroro}, which can be written as $Q_{r+1}
 \leq Q^2_r$, 
for $Q_r \leq 1$ (this last inequality,  starting from some r is a necessary
 condition of the convergence)  could be not satisfied unless
 both $ ||Q_{r+1}||$
 and $||Q_r||$ equal to
zero. This proves that stagnation of the iteration process is impossible and 
convergence of $||\delta u_{r+1}||$ to zero automatically leads to convergence
 of the QLM iteration sequence to the exact solution.
Hence the QLM assures not only convergence,
 but also convergence to the correct solution.

Another corollary of this iteration process
 is that if the solution and its derivatives are continuous
 functions of $x$,
the convergence of the QLM in the whole region will follow. Indeed, even if the zero
 iteration $u_0(x)$ is chosen not to satisfy
 the boundary conditions, the next iteration $u_1(x)$,  being a solution of
 a linear equation with linearized boundary conditions
 \ref{eq:bounconleft} and \ref{eq:bounconright}, will automatically satisfy the
 exact boundary conditions  \ref{eq:bounconl} and \ref{eq:bounconr}, at least up
 to the second order in difference
$\delta u_1$ at the boundaries. This means that the difference between 
the exact and
 first QLM iterations at some intervals near the boundaries will be small, so that
 the
QLM iterations in this interval would converge. Because the
 subsequent values of  $\it{k}\, \delta u_m(x), m>2$  became much 
smaller for this interval, in view of assumed continuity of the solution and its
 derivatives these differences will also be small at the neighboring intervals. The subsequent
 iterations will extend
the convergence to the next neighboring intervals and so on, until the convergence
 in the whole region will be reached. The predicted trend is therefore that
the QLM yields rapid convergence starting at the regions where the boundary conditions are imposed and then spreading from there to all other regions. 
 
An additional important corollary is that, in view of Eq. \ref{eq:unum},
 once the quasilinear iteration sequence
 starts to converge, it will continue to do so, unlike the
 perturbation expansion, which is often given by an asymptotic series and
  therefore converges only up to a certain order and  diverges thereafter.

Based on this summary of the QLM,  one can deduce the following important
 features of the quasilinearization method:

\begin{itemize}

\item[i)]
The method approximates the solution of nonlinear differential equations by
treating the nonlinear terms as a perturbation about the linear ones, and is
not based, unlike perturbation theories, on the existence of some kind of
small parameter.

\item[ii)]
The iterations converge uniformly and quadratically to the exact solution.
 In case of matrix $M_{st}$
 in Eq. \ref{eq:quadrilli} being strictly positive (negative) for all $x$
 in the interval, the convergence is also
 monotonic from below (above).
\item[iii)]
For rapid convergence it suffices  that an initial guess for
the zeroth iteration is sufficiently good to ensure the smallness of just
one of the quantities $q_r=\it{k}{\parallel} u_{r+1}-u_{r}{\parallel}$.
If the solution and its derivatives are continuous,   
 convergence follows from the fact that starting from the first iteration, 
all QLM iterations automatically satisfy the quasilinearized boundary
 conditions \ref{eq:bounconleft} and \ref{eq:bounconright}. The
convergence is extremely fast: if, for example, $q_1$ is of the order of
$\frac{1}{3}$, only 4 iterations are necessary to reach the accuracy of 8
digits, since $(\frac{1}{3})^{2^n}$ is of the order of
$(\frac{1}{10})^{2^{n-1}}$.

\item[iv)]
Convergence of $||\delta u_{r+1}||$ to zero automatically leads to convergence
 of the QLM iteration sequence to the exact solution.

\item[ v)]
Once the quasilinear iteration sequence at some interval starts to converge,
 it will always 
continue to do so. Unlike an asymptotic perturbation series, the
 quasilinearization method yield the required precision   
once a successful initial guess generates convergence after a few steps. 

\end{itemize}

\section{Analytically solvable example: comparison of quasilinearization
 approach with exact solution and with perturbation theory.}
In order to investigate the applicability of the quasilinearization
 method and its convergence and numerical stability, let us start from a simple
 example
 of an analytically solvable nonlinear ordinary differential equation suggested
 in ref. \cite{BMPS}:

\beq
u'(r)=- g\  u^n(r) ,  \ \ \  u(0)=1 ,  \label{eq:simple}
\eeq
where the boundary condition at $r=0$ is also given and  $'$ means
 differentiation in variable $r$. 
\nin
The exact solution to this problem is
\beq
u(r)= (1+(n-1)\, g\, r)^{-\frac{1}{n-1}} \label{eq:simplesol}
\eeq

\nin
Since
\beq
(1+x)^q=\sum_{0}^{\infty}\frac{\Gamma(q+1)}{m! \Gamma(q+1-m)}\ x^m ,  
\label{eq:bynom}
\eeq
\nin
the expansion of the solution \ref{eq:simplesol} in powers of $g$ is given by

\beq
u(r)=\sum_{0}^{\infty}\frac{\Gamma(\frac{n-2}{n-1}) (g(n-1))^m}{m! \Gamma(
\frac{n-2}{n-1}-m)}\; r^m 
 \label{eq:rbynom}
\eeq
\nin
The convergence radius of the series \ref{eq:rbynom} is
 $R=1/(g\, (n-1)),$ which is
 inversely proportional to the extent $n-1$ of the nonlinearity and to the value
 $g$ of the
 perturbation parameter.

Now consider the quasilinearization approach to this equation,
 taking, for example, $g=1$ and 
$n=6.$  Here we consider Eq. \ref{eq:simple} with a rather strong
 degree of nonlinearity.
 In this case,  one can expect the convergence of the perturbation expansion only
 up to 
$r\leq\ R = \frac{1}{5}$. 

The QLM procedure in the case where the nonlinear term  
depends only on the solution itself and not on its derivatives reduces to
setting $u_{k+1}'(r) = f(u_k) + (u_{k+1}(r) -u_{k}(r))\ f_u(u_k)$. Here 
$f= - g\; u^n(r) $ while its functional derivative $f_u$ equals 
to $ -g\ n u^{n-1}(r)$. 
The quasilinearized equation \ref{eq:difrec} for the $(k+1)$-th iteration for
 this case has therefore the following form:

\beq
u_{k+1}'(r) +n g\, u_k^{n-1}(r)\,  u_{k+1}(r)=(n-1)\,g\,u_k^n(r),
 \ \ \ \  u_{k+1}(0)=1\;, \label{eq:simplestar}
\eeq

\nin
where $u_k(r)$ is a previous iteration which is considered to be a known function.
Let us choose as a zero iteration $u_0(r)\equiv 1$ which satisfies the
 boundary condition $u_0(0)=1$. 

\begin{figure}
\begin{center}
\epsfig{file=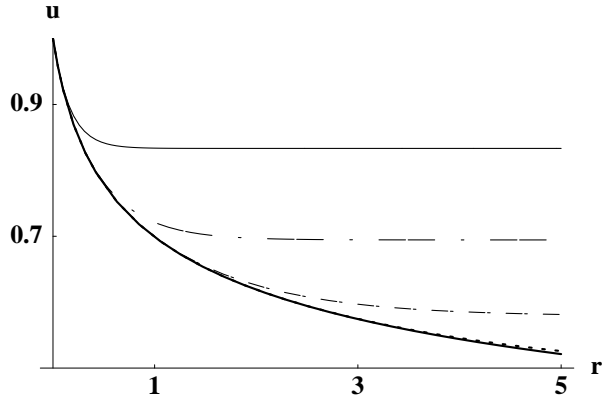,width=87mm}
\end{center}
\caption{Convergence of QLM iterations for the analytic example of
 section III and comparison with the exact solution. Thin solid,
dot-dashed, short-dashed and dotted curves correspond to the first, second,
third and fourth QLM iteration respectively, while the thick solid curve
 displays the exact solution. The convergence is monotonic from above
 as it should be according to the discussion in the text. The difference 
 between the exact solution and the sixth QLM iteration
 for all r in the figure is less than $ 10^{-6}$.}
\label{lqlmfig01}
\end{figure}

The results of our QLM calculations with Eq. \ref{eq:simplestar} are
 presented in Fig. \ref{lqlmfig01}
which displays the exact solution for the case of $n=6$ and $g=1,$
 together with
the first four QLM iterations. 
Convergence to the exact solution in Fig. \ref{lqlmfig01} is monotonic
 from above as it should be as discussed in Section II and in Refs.\cite{VBM,K59,BK65} due to fact that the
 second functional derivative
 $-n(n-1)u^{n-2}(x)$ of the left-hand side of Eq.\ref{eq:simple}
 for even $n$ is  strictly negative. The convergence starts at the boundary,
  exactly as  expected from the discussion in section II, 
and expands with each iteration to larger values of the variable $r$. 
The difference  between the exact solution and the sixth QLM iteration
 for all r in the  range between zero and five where our calculations
 were performed is less than $ 10^{-6}.$
  Note that the QLM yields a solution beyond the convergence radius 
limit
 on the series solution of $1/5.$ 

\section{Lane-Emden equation}

The Lane-Emden equation

\beq
y''(r)+\frac{2}{r}\; y'(r) +y^n(r)=0 ,\ \ \ \  y(0)=1,\; y'(0)=0 \label{eq:le}
\eeq

\nin
describes a variety of phenomena in theoretical physics and astrophysics, including aspects of 
stellar structure,  the thermal history of a spherical cloud of gas,  isothermal gas spheres,
and thermionic currents,  see ref. [17] and references therein.
The parameter $n$ defines an
 equation of state
with its physically interesting range being $0\leq n \leq 5$.  
The equation also appears in other contexts, e.g., in case of radiatively
 cooling,
 self gravitating gas clouds, in the mean-field treatment of a phase transition
 in critical absorption or in the modeling of clusters of galaxies.
For $n=0,1$ and $5$ the equation can be solved analytically. 
Setting $y=\frac{u}{r}$ transforms the equation to a more convenient form
without a first derivative:
\beq
u''(r) +\frac{u^n(r)}{r^{n-1}}=0 ,\; u(0)=0,\; u'(0)=1. \label{eq:lep}
\eeq

\nin
Let us consider this nonlinear equation for the
 physically interesting and analytically
 nonsolvable case of $n=4$. The quasilinearized form of
 equation \ref{eq:lep} is

\beq
u''_{k+1}(r) + n{\frac{u^{n-1}_k(r)}{r^{n-1}}}\;u_{k+1}(r)=\frac{n-1
}{r^{n-1}}\;u^n_k(r)
 ,\; u_{k+1}(0)=0,\; u'_{k+1}(0)=1. \label{eq:linlep}
\eeq

\begin{figure}
\begin{center}
\epsfig{file=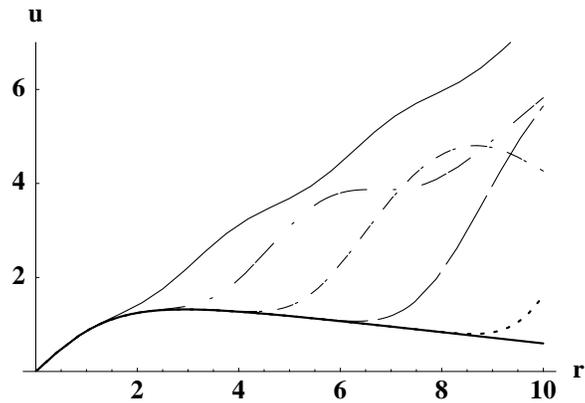,width=87mm}
\end{center}
\caption{Convergence of QLM iterations for the Lane-Emden
equation and comparison with the numerically obtained exact solution.
 Thin solid, dot-dashed, short-dashed, long-dashed and dotted curves
 correspond to the first, second, 
third, fourth  and fifth QLM iteration, respectively, while the
 thick solid curve displays the exact solution. The difference 
 between the exact solution and the eighth QLM iteration
 for all r in the figure is less than $ 10^{-11}$ .}
\label{lqlmfig02}
\end{figure}

\nin
The simplest initial guess, satisfying the boundary conditions
 will be $u_0(r)=r.$
 Comparison of the quasilinear solutions corresponding to the first
 five iterations with the numerically computed exact solution are given
 in Fig. \ref{lqlmfig02}. The figure shows that the convergence to
the exact solution
 is very fast. It starts, as in the example of the previous section, at the
 left boundary and spreads with each iteration to larger values of $r$
 as expected from the discussion in section II. The 
 difference  between the exact solution and the eighth QLM iteration
 for all $r$ in the  range between zero and ten,  where our calculations
 were performed,  is less than $ 10^{-11}$ .

\section{Thomas-Fermi equation}
The Thomas-Fermi equation \cite{T27,F28}

\beq
\sqrt{x}\  u''(x)=u^{\frac{3}{2}}(x) ,\ \ \ \ u(0)=1, \; u(\infty)=0,
 \label{eq:tfe}
\eeq

\nin
 is an equation for the electron density around the nucleus of the atom.
 The left hand side of the above equation equals zero for $u<0.$
  The Thomas-Fermi equation is also  very useful for
 calculating form-factors and for obtaining effective potentials
 which can be used as initial trial potentials in  self-consistent
 field calculations.. It is also applicable to the study of nucleons in nuclei
 and electrons in metal. 
It is long known (see \cite{BJ68} and references therein) that solution 
of this equation is very sensitive to a
value of the first derivative at zero which insures smooth and monotonic
 decay from $u(0)=1$ to $u(\infty)=0$ as demanded by boundary conditions.
  By contrast, the computation is much
 simpler for the quasilinearized version of this equation. The QLM procedure
 in this case reduces to
setting $u_{k+1}''(r) = f(u_k) + (u_{k+1}(r) -u_{k}(r))f_u(u_k)$,
where $f=\frac{u^{3/2}(r)}{\sqrt{x}} $ 
 and the functional derivative is $f_u = (3/2)\frac{u^{1/2}}{\sqrt{x}}$, so
 that
the QLM equation has a form:

\beq
\sqrt{x}\; u_{k+1}''(x)-\frac{3}{2}\; u_k^{\frac{1}{2}}(x)\; u_{k+1}(x) = 
-\frac{1}{2}\; u_k^{\frac{3}{2}}(x) ,\;\;u_{k+1}(0)=1, \;\; u_{k+1}(\infty)=0,
 \label{eq:tfequas}
\eeq

\begin{figure}
\begin{center}
\epsfig{file=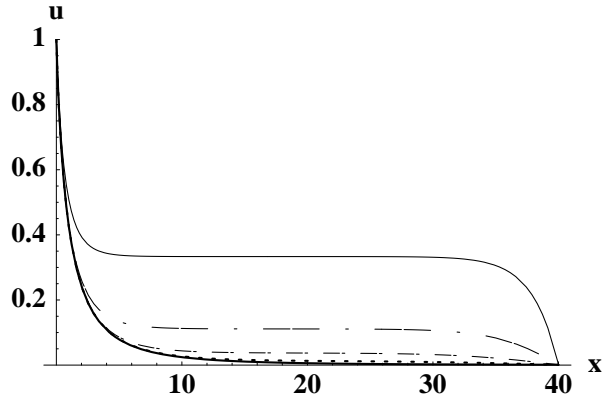,width=87mm}
\end{center}
\caption{Convergence of QLM iterations for the Thomas-Fermi
equation and comparison with the numerically obtained exact solution.
 Thin solid, dot-dashed, short-dashed and dotted curves correspond
 to the first, second, third and fourth QLM iteration,  respectively,
 while the thick solid curve displays the exact solution. The difference 
 between the exact solution and the eighth QLM iteration
 for all x in the figure is less than $ 10^{-7}$ .}
\label{lqlmfig03}
\end{figure}

\nin
 which is easily
 solved by specifying directly the boundary condition at infinity without
searching first for the proper value of the first derivative.
The initial guess, satisfying the boundary condition at zero was chosen to be
 $u_0(x)\equiv 1$.
The results of QLM calculations with Eq. \ref{eq:tfequas} are
 presented in Fig. \ref{lqlmfig03}
which displays the exact solution together with
 the first four QLM iterations. 
 The convergence starts at the boundaries,
  exactly as  expected from the discussion in section II, 
and expands with each iteration to a wider range of values of the variable $x$. 
The difference  between the exact solution and the eighth QLM iteration
 for all $x$ in the  range between zero and forty where our calculations
 were performed is less than $ 10^{-7}.$ 

\section{Classical anharmonic oscillator}

The classical anharmonic oscillator satisfies the nonlinear second-order Duffing equation

\beq
\ddot{u}(t) +u(t)+g\ u^3(t)=0  \label{eq:duffing}\ .
\eeq

\nin
The typical initial conditions are

\beq
u(0)=1,\;\; \dot{u}(0)=0 .  \label{eq:bounduffing}
\eeq
\nin
The solution oscillates strongly and thus is more difficult to approximate.
It is, for example, well known \cite{BMPS}  that the usual perturbative
 solution is valid
 only for times $t$ small compared with $\frac{1}{g}$, so that for
 larger $g$ the perturbative solution is adequate only on a small time
 interval.
  In contrast, the quasilinearization approach gives solution in the whole
 region also for large $g$-values. 

The quasilinearized equation is

\beq
\ddot{u}_{k+1}(t) +(1+3\ g\ u^2_k(t))u_{k+1}(t)-2\ g\  u_k^3(t)=0,\ \ \ 
 u_{k+1}(0)=1,\;\; \dot{u}_{k+1}(0)=0. \label{eq:qduffing}
\eeq 
\nin

\begin{figure}
\begin{center}
\epsfig{file=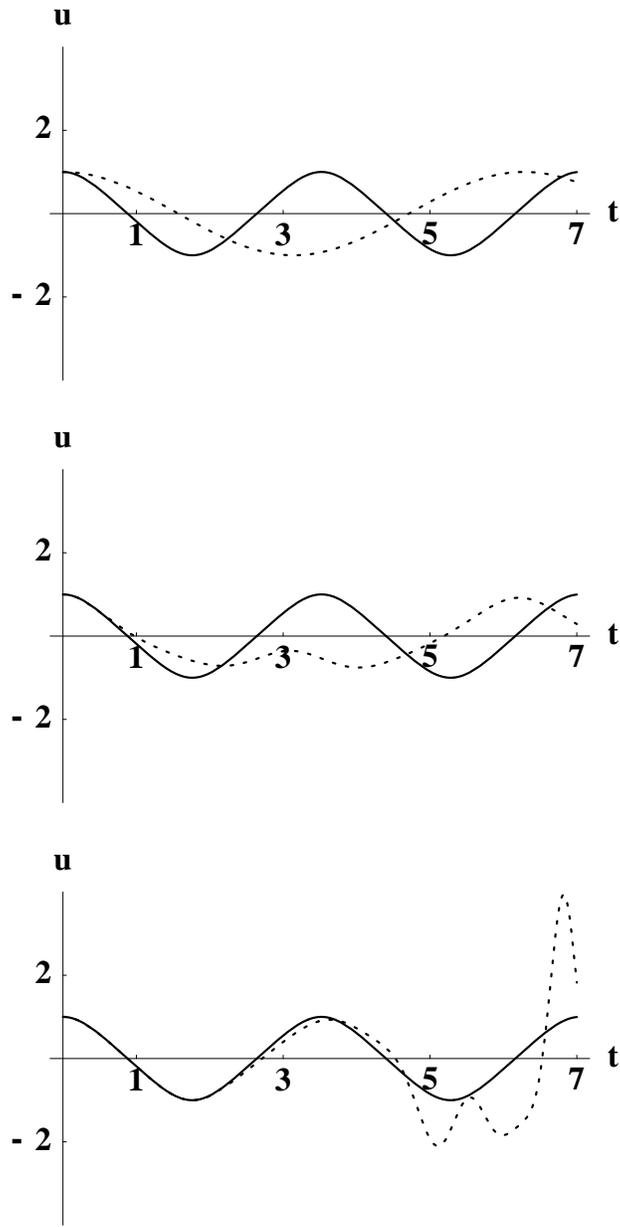,width=87mm}
\end{center}
\caption{Convergence of the first few QLM iterations for the Duffing
equation and comparison with the numerically obtained exact solution.
 The dotted curves on three consecutive graphs
 correspond to the first, second and fourth QLM iteration respectively, while the
solid curve displays the exact solution.}
\label{lqlmfig04}
\end{figure}

\begin{figure}
\begin{center}
\epsfig{file=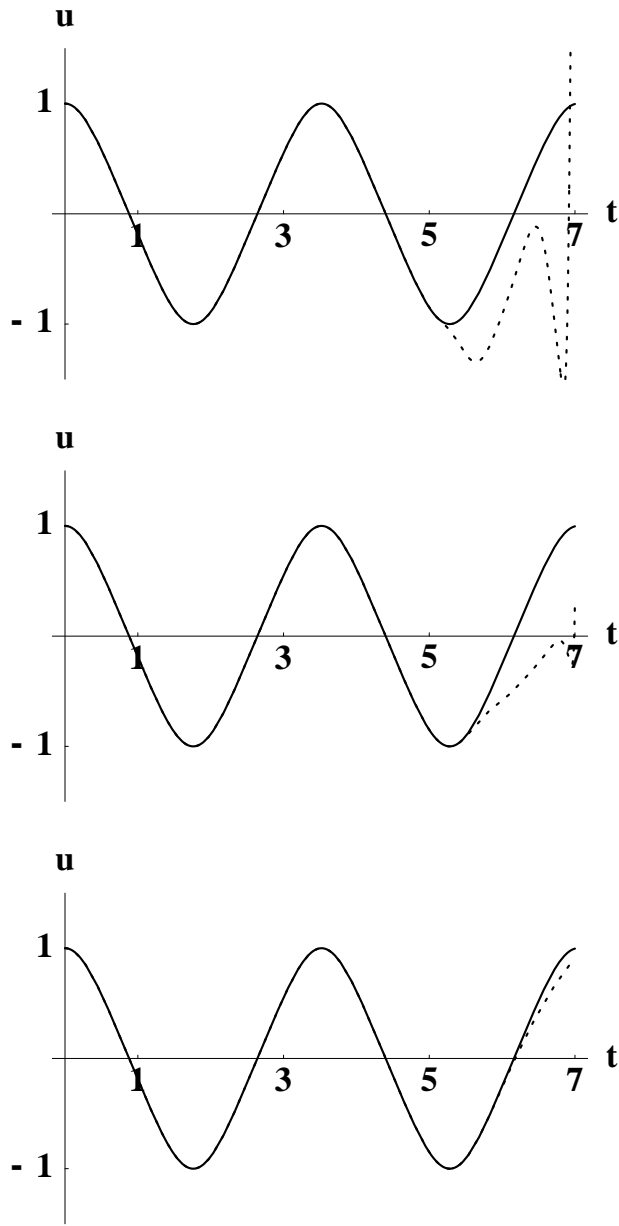,width=87mm}
\end{center}
\caption{Convergence of the higher QLM iterations for the Duffing
equation and comparison with the numerically obtained exact solution.
 The dotted curves on the three consecutive graphs
 correspond to the sixth,
seventh and eighth QLM iteration respectively, while the
solid curve displays the exact solution. The difference 
 between the exact solution and the eighth QLM iteration
 for all t in the figure is less than $ 10^{-10}$ .}
\label{lqlmfig05}
\end{figure}

The results of QLM calculations with Eq. \ref{eq:qduffing} for $g=3$ are
 presented in Figs. \ref{lqlmfig04} and \ref{lqlmfig05}.
Fig. \ref{lqlmfig04} displays the exact solution together with
the QLM solutions for the first, second and fourth iterations
while Fig. \ref{lqlmfig05} shows comparison of exact solution with 
sixth, seventh and eighth QLM iterations. 
Again, the convergence starts at the left boundary as  expected
 from the discussion in section II, 
and expands with each iteration to larger values of the variable $t$. 
The difference  between the exact solution and the eleventh QLM iteration
 for all t in the  range between zero and seven where
 our calculations were performed is less than $ 10^{-10}$ .

\section{Blasius equation}
A third order nonlinear Blasius equation  
\beq
u'''(x) +u''(x) u(x)=0 ,\ \ \  u(0)=u'(0)=0\;\; , u'(\infty)=1 \label{eq:blas}
\eeq
\nin
  describes the velocity profile of the fluid in the boundary layer. 
 The QLM procedure in this case is given by $u_{k+1}'''(x)=f(u_k,u_k'')+
(u_{k+1}-
u_k) f_u(u_k,u_k'')+(u_{k+1}''-u_k'') f_{u''}(u_k,u_k'')$, where
 $f(u,u')=-u''u$,  $f_u(u,u')=-u''$ and $f_{u''}(u,u')=-u$.
The quasilinearized version of the Blasius equation thus has a form

\begin{eqnarray}
u_{k+1}'''(x) +u_k(x) u_{k+1}''(x)+u_{k+1}(x) u_k''(x)-u_k(x) u_k''(x)=0 ,
\nonumber
\\ \;\; u_{k+1}(0)=u_{k+1}'(0)=0\;\; , u_{k+1}'(\infty)=1 . \label{eq:lianblas}
\end{eqnarray}
\nin

\begin{figure}
\begin{center}
\epsfig{file=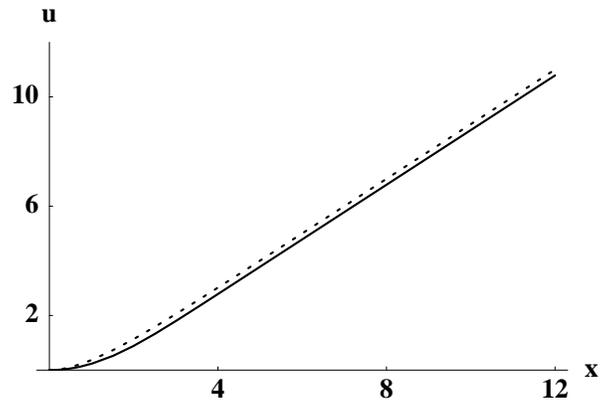,width=87mm}
\end{center}
\caption{Comparison of the first QLM iteration for the Blasius
equation with the numerically obtained exact solution. The difference 
 between the exact solution and the fifth QLM iteration
 for all x in the figure is less than $ 10^{-10}$ .}
\label{lqlmfig06}
\end{figure}

The initial guess, satisfying the boundary condition for the derivative 
at zero was chosen to be $u_0(x)\equiv 1$.
The results of QLM calculations with Eq. \ref{eq:lianblas} are
 presented in Fig. \ref{lqlmfig06} which displays the exact solution
 together with the first QLM iteration. The convergence starts at the
 left boundary as  follows from the discussion in section II, 
and expands with each iteration to larger values of the variable $x$. 
The difference  between the exact solution and the fifth QLM iteration
 for all $x$ in the  range between zero and ten where
 our calculations were performed is less than $ 10^{-11}$ .

\section{Conclusion}
Summing up, we formulated here the conditions under which the quadratic,
 uniform and often monotonic convergence of the quasilinearization
 method are valid.

We have followed here the derivation outlined in ref. \cite{VBM}, which is not
 based, unlike the derivations
 in refs.  \cite{K59,BK65}, on a smallness of the interval and on the boundness
of the nonlinear term and its functional derivatives, the conditions which
usually are not fulfilled in physical applications.

 In order to analyze and highlight the power and features of the
quasilinearization method (QLM), in this work we have also made numerical
computations on different ordinary  second and
 third order nonlinear differential equations, applied in physics, such as
  the Blasius, Duffing, Lane-Emden and Thomas-Fermi equations
 and have compared the results obtained by the quasilinearization
 method with the exact solutions. Although all our examples deal only with
 linear boundary conditions, the nonlinear boundary conditions can be
 handled readily after their quasilinearization as 
explained in Section II.

 In all the examples considered in the paper the simplest initial guess 
was enough to produce a quadratic convergence. However, as it is well known
for the Newton method on which quasilinearization method is based that
the divergence could occur if the initial guess is too bad. In this case
the modification of QLM based on the damped Newton method with relaxation
factor /cite {CB,RR}
 may help. The price of such modification would be, like in the
  damped Newton method only linear convergence instead of the quadratic one.

Our conclusions are as follows:

The QLM treats nonlinear terms  by a series of nonperturbative iterations
and is not based on the
existence of some kind of small parameter.  At every iterative stage,  the differential operator changes
significantly to account for the nonlinearity,  which is the major way that the QLM
differs from other approximative techniques. 
 As a result, as we see in all our examples, the QLM  is able to handle
  large values of the coupling constant and any degree of the nonlinearity,
 unlike  perturbation theory.  
 Thus the QLM provides extremely accurate and numerically
stable answers for a wide range of nonlinear physics problems. The QLM is also very easy to apply.

In view of all this, since most equations of physics, from classical
mechanics to quantum field theory, are either not linear or could be
transformed into a nonlinear form, the quasilinearization method  appears 
to be extremely useful and in many cases more advantageous than the
perturbation theory or its different modifications, like expansion in
inverse powers of the coupling constant, the ${1}/{N}$ expansion, etc.

\acknowledgments
The authors are indebted to Dr. Sch\"{o}nauer for careful reading
 of the manuscript and many constructive comments and suggestions.
The research was supported in part by the U.S. National Science Foundation
grant PHY-9970775 (FT) and by the Israel Science Foundation grant 131/00 (VBM).

\pagebreak

\newpage
\begin{center}
 FIGURE CAPTIONS
\end{center}

FIG.\ 1. {Convergence of QLM iterations for the analytic example of
 section III and comparison with the exact solution. Thin solid,
dot-dashed, short-dashed and dotted curves correspond to the first, second,
third and fourth QLM iteration respectively, while the thick solid curve
 displays the exact solution. The convergence is monotonic from above
 as it should be according to the discussion in the text. The difference 
 between the exact solution and the sixth QLM iteration
 for all r in the figure is less than $ 10^{-6}$.}

\bigskip

FIG.\ 2. {Convergence of QLM iterations for the Lane-Emden
equation and comparison with the numerically obtained exact solution.
 Thin solid, dot-dashed, short-dashed, long-dashed and dotted curves
 correspond to the first, second, 
third, fourth  and fifth QLM iteration, respectively, while the
 thick solid curve displays the exact solution. The difference 
 between the exact solution and the eighth QLM iteration
 for all r in the figure is less than $ 10^{-11}$ }

\bigskip

FIG.\ 3. {Convergence of QLM iterations for the Thomas-Fermi
equation and comparison with the numerically obtained exact solution.
 Thin solid, dot-dashed, short-dashed and dotted curves correspond
 to the first, second, third and fourth QLM iteration,  respectively,
 while the thick solid curve displays the exact solution. The difference 
 between the exact solution and the eighth QLM iteration
 for all x in the figure is less than $ 10^{-7}$ .}

\bigskip

FIG.\ 4. {Convergence of the first few QLM iterations for the Duffing
equation and comparison with the numerically obtained exact solution.
 The dotted curves on three consecutive graphs
 correspond to the first, second and fourth QLM iteration respectively, while the
solid curve displays the exact solution.}

\bigskip

FIG.\ 5. {Convergence of the higher QLM iterations for the Duffing
equation and comparison with the numerically obtained exact solution.
 The dotted curves on the three consecutive graphs
 correspond to the sixth,
seventh and eighth QLM iteration respectively, while the
solid curve displays the exact solution. The difference 
 between the exact solution and the eighth QLM iteration
 for all t in the figure is less than $ 10^{-10}$ .}

\bigskip

FIG.\ 6. {Comparison of the first QLM iteration for the Blasius
equation with the numerically obtained exact solution. The difference 
 between the exact solution and the fifth QLM iteration
 for all x in the figure is less than $ 10^{-10}$ .}


\end{document}